# Superconducting and Topological Properties in Centrosymmetric PbTaS$_2$ Single Crystals

J. J. Gao[1, 2,†], J. G. Si[1, 2,†], X. Luo[1*], J. Yan[1, 2], Z. Z. Jiang[1, 2], W. Wang[1, 2], C. Q. Xu[5], X. F. Xu[5], P. Tong[1], W. H. Song[1], X. B. Zhu[1], W. J. Lu[1*], and Y. P. Sun[3, 1, 4*]

[1] Key Laboratory of Materials Physics, Institute of Solid State Physics, Chinese Academy of Sciences, Hefei, 230031, China

[2] Science Island Branch of Graduate School, University of Science and Technology of China, Hefei, 230026, China

[3] Anhui Province Key Laboratory of Condensed Matter Physics at Extreme Conditions, High Magnetic Field Laboratory, Chinese Academy of Sciences, Hefei, 230031, China

[4] Collaborative Innovation Center of Advanced Microstructures, Nanjing University, Nanjing, 210093, China

[5] Department of Physics, Changshu Institute of Technology, Changshu, 215500, China

**ABSTRACT:** We report the superconductivity of $PbTaS_2$ single crystals with the centrosymmetric structure. The systematic measurements of magnetization, electric transport and specific heat indicate that $PbTaS_2$ is a weakly coupled type-II superconductor with transition temperature $T_c \sim 2.6$ K. Furthermore, the band structure calculations predicted four nodal lines near the Fermi energy with 'drumhead-like' surface states, suggesting centrosymmetric $PbTaS_2$ is a candidate of topological nodal line semimetals. These results demonstrate that $PbTaS_2$ may open up another avenue for further exploring the properties of superconductivity and topological nodal-line states.

# INTRODUCTION

The discovery of topological insulators has stimulated broad research interests in exploring–novel topological states and materials.[1-4] As Weyl semimetals observed in the experiments, the research interest in topological phenomena has gradually expanded from insulators to semimetals.[5-9] While topological semimetals are a new type of materials with one-dimensional Fermi lines or zero-dimensional Weyl or Dirac points in momentum space,[10,11] which can be further classified into Dirac semimetal (protected by certain crystalline symmetry),[12,13] Weyl semimetal (protected by lattice translation)[5,14] and nodal line semimetal.[15] Among these topological materials, some of them exhibit superconductivity under charge carrier doping or external pressure conditions,[16-18] and some are even superconductors themselves.[10,19,20] Therefore, it is of interest to determine whether there is any interplay between the topological properties and the superconductivity. What's more, the search for topological superconductors (TSCs) has become one of the most urgent subjects in condensed matter physics.

Recently, the noncentrosymmetric superconductor $PbTaSe_2$ with space group of $P\bar{6}m2$ (No.187) has been the focus of much interest. Based on the detailed angle-resolved photoemission measurements (ARPES) and systematic theoretical study, the existence of topological nodal-line (NL) states in $PbTaSe_2$ was confirmed by Bian *et al.*.[10] Using quasi-particle scattering interference imaging, Guan *et al.* reported two topological surface states (TSSs) with a Dirac point at E ≈ 1.0 eV in $PbTaSe_2$, of which the inner TSS and the partial outer TSS cross $E_F$, on the

Pb-terminated surface of this fully gapped superconductor.[21] These results suggest that PbTaSe$_2$ is a potential candidate for TSC. Similar to PbTaSe$_2$, theoretical research indicates that the noncentrosymmetric PbTaS$_2$ (No.187) is also a topological NL semimetal.[11] Although PbTaS$_2$ is also a superconductor reported as early as 1973,[22] it's noting that PbTaS$_2$ mentioned here has the centrosymmetric structure with space group of P6$_3$/mmc (No.194), which is quite different from PbTaSe$_2$. Then, in which structure PbTaS$_2$ could exist stably and whether the superconducting and topological properties could be found in this stable structure are still puzzling. In order to get a deeper insight into the basic properties of PbTaS$_2$, further research is needed. In this work, we have successfully grown the single crystals of PbTaS$_2$, which crystallizes in the P6$_3$/mmc space group. Based on the magnetization, electric transport and specific heat measurements in detail, the superconductivity of centrosymmetric PbTaS$_2$ can be determined, which is confirmed by the theoretical calculations. In addition, the first-principles calculations show that centrosymmetric PbTaS$_2$ is also a topological NL semimetal.

## EXPERIMENTAL SECTION AND CALCULATIONS

**Sample Preparation.** PbTaS$_2$ single crystals were grown by the chemical vapor transport (CVT) method with lead chloride as the transport agent. Polycrystalline PbTaS$_2$ was synthesized previously by the solid-state reaction of Pb, Ta and S powders in an evacuated quartz tube at 900 °C. Then, the obtained materials were ground, sealed in an evacuated quartz tube with PbCl$_2$, and heated in a two-zone furnace with 850 °C in hot zone and 800 °C in cold zone for 10 days.

**Basic properties measurements.** Single crystal and powder X-ray diffraction (XRD) experiments were performed by the PANalytical X'pert diffractometer using the Cu $K_{\alpha 1}$ radiation ($\lambda$ = 0.15406 nm) at room temperature. Electrical transport and specific heat measurements were carried out in a Quantum Design Physical Property Measurement System (PPMS). Magnetic properties were performed by the Magnetic Property Measurement System (MPMS).

**Computational Details.** The first-principles calculations were performed by Quantum Espresso (QE) code.[23,24] The pseudopotentials we used were ultrasoft pseudopotentials according to the generalized gradient approximation (GGA), parameterized by Perdew-Burke-Ernzerhof (PBE).[25] The energy cutoff of 800 Ry (80 Ry) was chosen for the charge density (wave functions) basis. The Brillouin zone (BZ) was sampled by a 16 × 16 × 4 $k$-points grid, ensuring that convergence criterion for energy (force) is less than $10^{-8}$ Ry ($5\times10^{-4}$ Ry/Å). For electron-phonon coupling calculations, a denser 32 × 32 × 8 $k$-points grid and an 8 × 8 × 2 $q$-points grid were used. To investigate the projected surface states, we constructed first-principles tight-binding model Hamiltonian by using maximally localized Wannier function (MLWF) method,[26,27] where the $d$ orbits of Ta atom and the $p$ orbits of the Pb and S atoms were used as basis. Then, we used Wannier Tools package and the constructed Hamiltonian to compute the surface states and the isoenergy contour.[28]

## RESULTS AND DISCUSSION

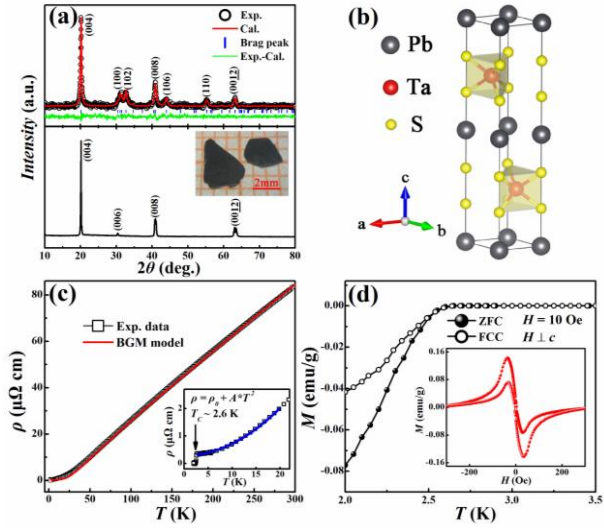

**Figure 1.** The top of (a) is the powder XRD pattern of the crushed PbTaS$_2$ crystals measured at room temperature. XRD pattern of the single crystal with (00$l$) reflections is shown at the bottom of (a), inset presents the picture of the studied PbTaS$_2$ single crystal. (b) The crystal structure of centrosymmetric PbTaS$_2$. (c) Resistivity as a function of the temperature with current flowing in the ab plane. The red solid line is fitting by using the Bloch–Grüneisun–Mott (BGM) model. The inset presents an enlarged view of the plot in a low temperature range ($T < 25$ K), the blue solid line is the fitting to the data obtained from the Fermi-liquid model. (d) The zero-field-cooling (ZFC) and field-cooling (FC) magnetic susceptibility curve measured with magnetic field of 10 Oe and perpendicular to the c axis. The inset of (d) shows the magnetic field dependence of magnetization at $T = 2$ K.

The top of Figure 1(a) presents the powder XRD pattern. The lattice parameters are refined and listed in **Table S1**. According to the fitting results of the powder XRD, we could determine that PbTaS$_2$ crystallizes in a layered hexagonal structure with the

P6$_3$/mmc space group, which demonstrates inversion symmetry. At the bottom of Figure 1(a), the single-crystal XRD pattern recorded (00*l*) planes of preferred orientation for PbTaS$_2$, and the inset presents the picture of the studied PbTaS$_2$ single crystal. As shown in Figure 1(b), the layered structure of centrosymmetric PbTaS$_2$ is formed by the alternative stacking of TaS$_2$ and Pb layers. The lattice can also be viewed as a Pb layer intercalating two adjacent TaS$_2$ layers, where Ta atoms are located at unequal positions. Based on this structure, the phonon dispersion curves were calculated, as shown in Figure S2. We found that no imaginary frequencies in whole Brillouin Zone, which suggests the dynamic stability of the considered crystal structure. Therefore, it should be noted that PbTaS$_2$ preserves the inversion symmetry, which is quite different from PbTaSe$_2$. Figure 1(c) displays the temperature dependence of resistivity of PbTaS$_2$ single crystals from 300 K to 2 K with current applied in the *ab* plane. The resistivity data shows a metallic behavior with the large residual resistivity ratio (RRR) ~ 251, which indicates the high quality of the sample. The $\rho(T)$ curve was fitted to the Bloch-Grüneisen-Mott (BGM) model over the whole temperature range.[29,30] From the inset of Figure 1(c), the superconducting transition ($T_c$) can be observed at about 2.6 K. We further applied the Fermi-liquid model $\rho(T) = \rho_0 + AT^2$ to the $\rho(T)$ curve below 20 K, where $\rho_0$ and A are the residual resistivity and a constant, respectively. The model analysis indicates the Fermi-liquid-like behavior for PbTaS$_2$ single crystals and the fitting parameters were listed in **Table 1**. Figure 1(d) presents the temperature-dependent magnetization in the zero-field and field cooling (ZFC and FC) modes with magnetic field of 10 Oe parallel to the *ab* plane. The $T_c$ determined from the susceptibility curve is about 2.6 K, which is consistent with the resistivity result. The magnetization as a function of magnetic fields at 2 K is shown in the inset of Figure 1(d), and the clear hysteresis can be observed, which suggests a typical type-II superconducting behavior of PbTaS$_2$.

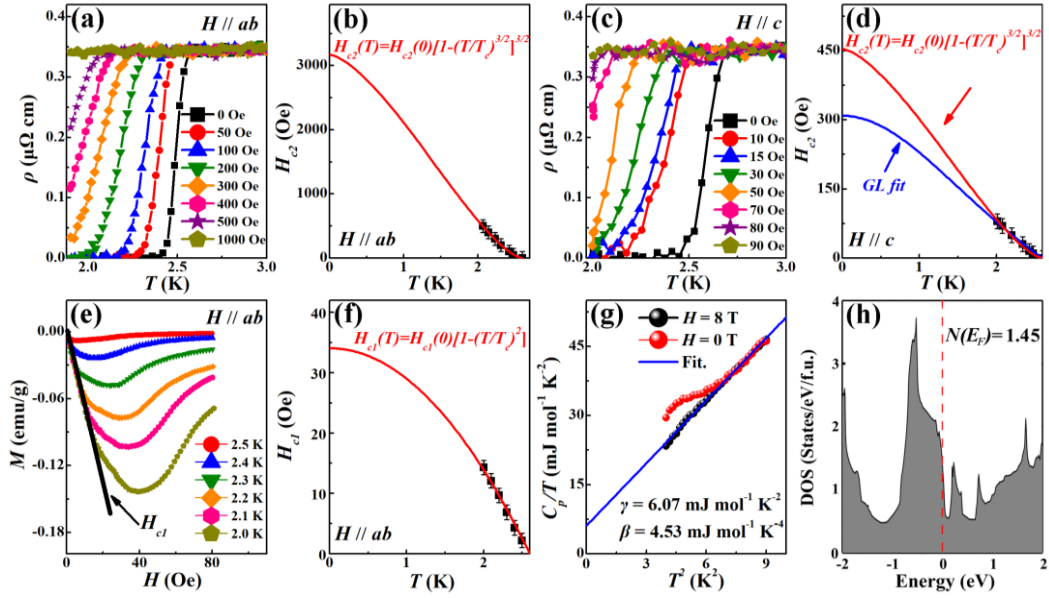

**Figure 2.** The low temperature in-plane resistivity $\rho(T)$ curves under different magnetic fields for (a) $H$ // ab and (c) $H$ // c. The temperature dependence of the upper critical field extracted from the $\rho(T)$ curves with (b) $H$ // ab and (d) $H$ // c. (e) The magnetic field dependence of magnetization at different temperatures and (f) temperature dependence of the lower critical field, with $H$ parallel to the ab plane. (g) $C_p/T$ as the function of $T^2$ under the magnetic fields of 0T (red solid symbols) and 8 T (black solid symbols). The curve for 8 T is fitted by the relation $C_p/T = \gamma_n + \beta T^2$. (h) The density of states (DOS) of PbTaS$_2$.

We also performed the measurements of the magnetic field-dependent magnetization ($M(H)$) at various temperatures. In the case of PbTaS$_2$, the $M(H)$ curves are quite different with $H // ab$ or $H // c$, which means the anisotropic of lower critical field ($H_{c1}$). It is worth noting that the demagnetization effect cannot be ignored. However, it is easy for us to determine the $H_{c1}$ for $H // ab$ ($H_{c1}^{ab}$) from $M(H)$ curves directly with considering the demagnetization factor (N) ~ 0, as shown in Figure 2(e). Conversely, for the case of $H // c$, the demagnetization effect (N ~ 1) would make a higher field that the sample feels around it. Therefore, it is difficult for us to determine the $H_{c1}$ for $H // c$ ($H_{c1}^{c}$) from $M(H)$ curves. Figure 2(f) presents $H_{c1}$ as a function of $T$, by fitting the experimental data to the formula $H_{c1}(T) = H_{c1}(0)[1-(T/T_c)^2]$, we can estimate $H_{c1}(0)$. The lower critical field at zero temperature for $H // ab$ is $H_{c1}^{ab}(0) = 34.06$ Oe. The low temperature in-plane resistivity for PbTaS$_2$ under two directions of various magnetic fields with $H // ab$ and $H // c$ are presented in Figure 2(a) and (c), respectively. We choose $T_c^{onset}$ as the critical temperature and plot the upper critical field values vs temperature for $H // ab$ and $H // c$ in Figure 2(b) and (d). And the whole $H_{c2}(T)$ data can be fitted by using the previously proposed formula $H_{c2}(T) = H_{c2}(0)(1-t^{3/2})^{3/2}$, where $t = T/T_c$. The value of the upper fields are about $H_{c2}^{ab}(0)$ ~ 3171.32 Oe and $H_{c2}^{c}(0)$ ~ 458.86 Oe. We found that $\mu_0 H_{c2}(0)$ is much lower than the Pauli paramagnetic limit field, $\mu_0 H_P = 1.86 T_c = 4.84$ T, which indicating the upper critical field is limited by the orbital effect. Based on the Ginzburg-Landau theory, we could estimate the superconducting coherent length $\xi$ by using the formulas: $H_{c2}^{c}(0) = \phi_0/(2\pi\xi_{ab}^2(0))$ and $H_{c2}^{ab}(0) = \phi_0/(2\pi\xi_{ab}(0)\xi_c(0))$, where $\phi_0 = h/2e$ is the magnetic flux quantum.[31] And the GL parameter $\kappa(0)$ along different magnetic field directions can also be obtained by the equation $H_{c2}^{ab}(0)/H_{c1}^{ab}(0) = 2\kappa_{ab}^2(0)/\ln\kappa_{ab}(0)$ (or

$H_{c2}{}^c(0)/H_{c1}{}^c(0) = 2\kappa_c{}^2(0)/\ln\kappa_c(0))$. What's more, according to the relationships between the GL parameter $\kappa(0)$, coherent length $\xi(0)$ and penetration length $\lambda(0)$: $\kappa_c(0) = \lambda_{ab}(0)/\xi_{ab}(0)$ and $\kappa_{ab}(0) = \lambda_{ab}(0)/\xi_c(0)$, we can determine the related parameters (even estimate $H_{c1}{}^c(0)$) as listed in **Table 1.**

**Table 1.** The superconducting parameters of PbTaS$_2$ single crystal.

| Parameters | Units | $H \parallel ab$ | $H \parallel c$ |
|---|---|---|---|
| $T_c$ | K | 2.6 | |
| $\rho_0$ | $\mu\Omega$ cm | 0.29 | |
| $A$ | $\mu\Omega$ cm K$^{-2}$ | 0.00432 | |
| RRR | | 251 | |
| $H_{c1}(0)$ | Oe | 34.06 | 40.76 (estimated) |
| $H_{c2}(0)$ | Oe | 3171.32 | 458.86 |
| $\xi(0)$ | Nm | 85.14 | 12.18 |
| $\lambda(0)$ | Nm | 127.28 | 889.71 |
| $\kappa(0)$ | | 10.45 | 1.49 |
| $\gamma_n$ | mJ mol$^{-1}$ K$^{-2}$ | 6.07 | |
| $\beta$ | mJ mol$^{-1}$ K$^{-4}$ | 4.53 | |
| $\Theta_D$ | K | 120 | |
| $\lambda_{ep}$ | | 0.69 | |

Figure 2(g) shows the temperature dependence of the specific heat $C_p$ from 2 K to 120 K, and the inset (right) illustrates $C_p/T$ versus $T^2$ low-temperature range measured under the magnetic fields of $\mu_0H = 0$ and 8 T. At $\mu_0H = 0$ T, the specific heat displays the superconducting feature as indicated by the jump at about 2.4 K, which is completely suppressed under the magnetic field of 8 T. The experimental data can be well fitted by using the formula $C_p/T = \gamma_n + \beta T^2$. The fitting result yields the normal state Sommerfeld coefficient $\gamma_n$ = 6.07 mJ mol$^{-1}$ K$^{-2}$, and the phonon specific coefficient $\beta$ = 4.53 mJ mol$^{-1}$ K$^{-4}$. The Debye temperature $\Theta_D$ is equal to 120 K, determined from the formula $\Theta_D = [12\pi^4 nR/(5\beta)]^{1/3}$. Furthermore, we calculated the electron-phonon coupling constant $\lambda_{ep}$ by using the McMillans formula:[32]

$$\lambda ep = \frac{1.04 + \mu^* \ln(\frac{\Theta_D}{1.45T_c})}{(1 - 0.62\mu^*)\ln(\frac{\Theta_D}{1.45T_c}) - 1.04}.$$

Assuming the Coulomb pseudopotential $\mu^*$ = 0.13 (the value has been widely used in similar materials, like PbTaSe$_2$ and SnTaS$_2$),[19,20] $\lambda_{ep}$ is found to be 0.69, which is close to the calculated value 0.71 (the calculation process can be found in the Supporting Information). The value supports the weakly coupling scenario. With the results mentioned above, we can also calculate the noninteracting density of states at the Fermi energy from $N(E_F) = 3\gamma / [\pi^2 k_B^2(1 + \lambda_{ep})]$, which gives $N(E_F) \sim$ 1.54 states eV$^{-1}$ per formula unit. The value is quite close to the 1.45 states eV$^{-1}$ per formula unit calculated from theoretical predictions, as shown in Figure 2(h).

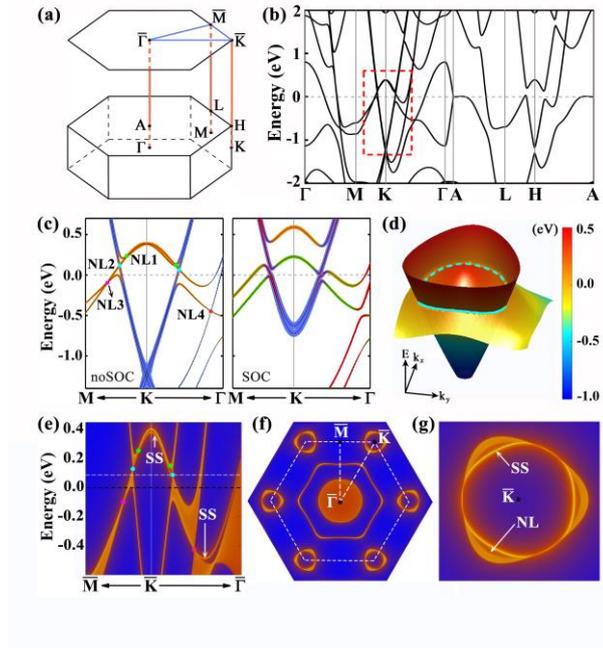

**Figure 3.** (a) Bulk and (001) surface Brillouin zone of PbTaS$_2$. (b) Calculated band structure of PbTaS$_2$ without SOC. (c) Zoom-in band structures around K without/with SOC. The colour code shows the orbital components (Orange and blue are corresponding to Ta-$d_{xy}$/$d_{x2-y2}$ and Pb-$p_x$/$p_y$ respectively without SOC, while considering SOC, the orbitals would split into spin up (orange and blue) and spin down (green and red) along the z axis). The size of the band is proportional to the weight of projection onto atomic orbitals. (d) Three dimensional band structure on $k_x$-$k_y$ plane of $k_z = 0$. The NL2 is highlight by the light blue. (e) Surface band structure of PbTaS$_2$ along $\overline{M}$-$\overline{K}$-$\overline{\Gamma}$. The surface states are pointed by white arrows. (f) The isoenergy contour showing the NL and SS states, which can be seen clearly in (g).

As mentioned above, PbTaS$_2$ has many similarities with the topological NL semimetal PbTaSe$_2$, we have also investigated the electronic structures of PbTaS$_2$ based on the density functional theory (DFT). The Brillouin zones of the bulk and

(001)-projected surface are shown in Figure 3(a). Figure 3(b) presents a whole band structure of PbTaS$_2$ without spin-orbit coupling (SOC), showing a clear metallic character with several bands crossing the Fermi level and a giant hole pocket around Γ point. Close to the Fermi energy, the band structure is mainly contributed by Ta-$d_{xy}/d_{x^2-y^2}$ and Pb-$p_x/p_y$ (see in Figure S3). The zoom-in views of the band structure around K point without and with SOC are shown in Figure 3(c). In the absence of SOC, there exit six band crossing points along M - K and K - Γ directions and band inversion between Ta-$d_{xy}/d_{x^2-y^2}$ and Pb-$p_x/p_y$. We perform a more accurate band calculation and find the band crossing points around the high symmetry point of K are not isolated but forming four NLs in the plane of $k_z = 0$. We plot the band structure with considering the symmetry, and find the band crossing points of NL3 and NL4 belong to two pairs of different bands (see in Figure S4). which is inconsistent with the previous report.[20] At present, the NL semimetals have been mainly divided into three types:[33] type A has mirror reflection symmetry,[34,35] type B is with the coexistence of time-reversal symmetry and space inversion symmetry,[36-40] and type C has the nonsymmorphic space group with glide plane or screw axes symmetries.[41-43] All these types are classified according to the symmetry, therefore, it is obvious that PbTaS$_2$ belongs to the type B due to retaining the time-reversal and space inversion symmetry. Similar to Weyl/Dirac semimetals, NL semimetals could also be classified into three types based on the dispersion of points on the NL: type I, type II and hybrid NL (composed of both type I and type II points), of which type I has been proposed and realized in many compounds.[44-47] While the type-II NL has also been observed in

Mg$_3$Bi$_2$ by ARPES recently.[48] Typically, NL2 belongs to the type I, and the three-dimensional (3D) schematic of it can be seen in Figure 3(d). While the other NLs are shown in Figure S5, and the classifications of them are more complicated. When the SOC is turned on, the band splits into two band branches with opposite spin orientations and the NLs are gapped, as indicated at the right of Figure 3(c). Since NLs originate from multiple bands crossing, it is not convenient to evaluate the topology of PbTaS$_2$, which is quite different from the case of PbTaSe$_2$.[49] The size of SOC-induced gaps is in the range of 44-300 meV (the results of band gap corrected by the typical screened hybrid HSE approximation (HSE06) can also be found in Figure S6 and **Table S2**). Especially, the size of SOC-induced gap nearest the Fermi energy (NL2) in PbTaS$_2$ is just about 40 meV. The value is comparable with those in typical NL materials, like ZrB$_2$ (~ 40 meV), the NL signature in which has been experimentally verified by ARPES.[50] An important characteristic of the topological NL semimetals is the existence of 'drumhead-like' surface states. We calculated the (001)-projected surface states around K-point as shown in Figure 3(e). One can see that four NLs highlighted by four different color balls and the 'drumhead-like' surface states represented by the arrows. In order to get the comprehensive view of the node ring and 'drumhead-like' surface states, the isoenergy contour was calculated and presented in Figure 3(f) and (g). The energy of the isoenergy contour is set as E = 0.16 eV in the vicinity of the NL2 and NL1, as indicated by the white dash line in Figure 3(e). The 'drumhead-like' surface state and the NL are obviously presented in Figure 3(g). Comparing with the NLs protected by mirror reflection symmetry in PbTaSe$_2$, the NLs in PbTaS$_2$ are protected by time-reversal and inversion symmetries, which is consistent with SnTaS$_2$ reported recently.[20] These results may pave the way for exploring the different origins of topological NL states.

## CONCLUSIONS

In summary, we have grown the single crystals of PbTaS$_2$ with the centrosymmetric structure, the superconductivity of which could be determined by the systematic measurements of magnetization, electric transport and specific heat. The resistivity and magnetic susceptibility indicate that PbTaS$_2$ is a type-II superconductor, while the value of $\lambda_{ep}$ deduced from the specific heat curves suggest it fits the weakly coupling scenario, which is consistent with the electron-phonon coupling calculations. The band structure calculation shows four NLs in the $k_z = 0$ plane near the Fermi energy with 'drumhead-like' surface states without considering spin-orbit coupling. All the results demonstrate that the centrosymmetric superconductor PbTaS$_2$ is a promising candidate for TSC researches, which may open up another avenue for further exploring the properties of superconductivity and topological NL states.

■ **ASSOCIATED CONTENT**

**Supporting Information**

The computational and experimental details, magnetic data and some extended results of theoretical calculations.

## Author Contributions

[†]These authors contributed equally to this work.

## Notes

The authors declare no competing financial interests.

## ■ ACKNOWLEDGMENTS

This work was supported by the National Key Research and Development Program under Contracts 2016YFA0300404 and the National Nature Science Foundation of China under Contracts 11674326, 11874357, 11774351, 11974061, the Joint Funds of the National Natural Science Foundation of China and the Chinese Academy of Sciences' Large-Scale Scientific Facility under Contracts U1832141, U1932217 and the Key Research Program of Frontier Sciences, CAS (QYZDB-SSW-SLH015) and The uses with Excellence and Scientific Research Grant of Hefei Science Center of CAS (2018HSC-UE011).